# The VORTEX coronagraphic test bench

A. Jolivet*[a], P. Piron[a], E. Huby[a], O. Absil[a], C. Delacroix[a], D. Mawet[b] and S. Habraken[a]
[a]Department of Astrophysics, Geophysics and Oceanography, University of Liège,
17 allée du Six Août, B-4000 Sart Tilman, Belgium
[b]European Southern Observatory, Alonso de Cordóva 3107, Vitacura, Santiago, Chile

## ABSTRACT

In this paper, we present the infrared coronagraphic test bench of the University of Liège named VODCA (Vortex Optical Demonstrator for Coronagraphic Applications). The goal of the bench is to assess the performances of the Annular Groove Phase Masks (AGPMs) at near- to mid-infrared wavelengths. The AGPM is a subwavelength grating vortex coronagraph of charge two (SGVC2) made out of diamond. The bench is designed to be completely achromatic and will be composed of a super continuum laser source emitting in the near to mid-infrared, several parabolas, diaphragms and an infrared camera. This way, we will be able to test the different AGPMs in the M, L, K and H bands. Eventually, the bench will also allow the computation of the incident wavefront aberrations on the coronagraph. A reflective Lyot stop will send most of the stellar light to a second camera to perform low-order wavefront sensing. This second system coupled with a deformable mirror will allow the correction of the wavefront aberrations. We also aim to test other pre- and/or post-coronagraphic concepts such as optimal apodization.

## 1. INTRODUCTION

We are developing at the University of Liège (Belgium) a vortex coronagraph based on subwavelength gratings (SGVC). The goal of coronagraphy is to observe faint stellar companions or exoplanets around bright stars by cancelling the on-axis starlight. The first generation of these SGVCs is known as the Annular Groove Phase Mask (AGPM, Mawet et al. 2005. [1]), and shows very promising results. Three of them have now been installed on world-leading diffraction-limited infrared cameras, namely VLT/NACO, VLT/VISIR and LBT/LMIRCam. As a part of the VORTEX project, it has become essential to have our own dedicated optical bench at the University of Liège to perform the tests and to assess the quality of the produced components.

In this manuscript, we present the Vortex Optical Demonstrator for Coronagraphic Applications (VODCA), an optical test bench designed in a first step to assess the performances of the AGPM at near- to mid-infrared wavelengths. In a second step we will test pre- and post- coronagraphic concepts to improve starlight cancellation.

*ajolivet@ulg.ac.be; phone +3243663721; ago.ulg.ac.be

## 2. VODCA REQUIREMENTS

The AGPM is a subwavelength grating vortex coronagraph of charge two (SGVC2) made out of diamond (see Fig. 1). VODCA is designed to assess the coronagraphic performance of an SGVC, in other words its ability to cancel the starlight when used at the telescope. Our first recently manufactured AGPMs are designed to operate in the near- to mid- infrared. VODCA will have to work in this range of wavelength. Typically we want to cover the H, K, L and M bands. Covering the N band where the AGPM also operates, would require a fully cryogenic bench, which is beyond the scope of this project.

AGPMs are achromatic on broad band thanks to the subwavelength grating design. In order to completely characterize the starlight rejection, we will use a continuum spectrum infrared light to match the operating conditions on a telescope as close as possible. Under these conditions, the difficulty is to design a bench completely achromatic from 1 to 5 µm.

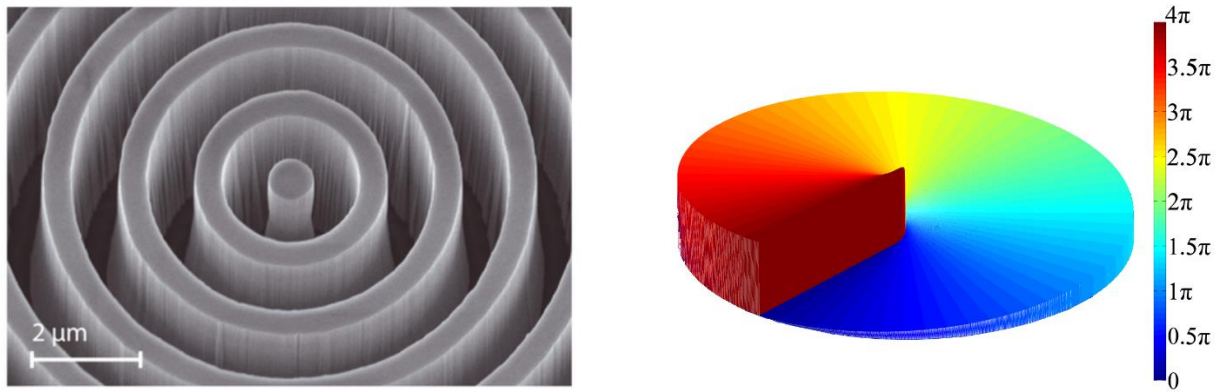

Figure 1. Left: Scanning electron microscope (SEM) picture of the center of an annular groove phase mask (AGPM) made out of diamond and dedicated to coronagraphic applications in the L band - Delacroix 2013.

Right: Phase ramp profile around the center of an AGPM (charge 2) – Piron 2014.

## 3. DESCRIPTION OF THE HARDWARE

### 3.1 Optics

Taking into account all the considerations presented in Sect. 2, the final design must include optics generating the lowest chromatic aberrations. Hence, we chose to use only mirrors on the bench.

We select two parabolic mirrors with a diameter of 152.4 mm and an effective focal length of 609.6 mm. Both of them have a protected aluminum coating ensuring an excellent reflectivity in the near- and mid- infrared, the range of wavelength where VODCA operates. The parabolas are used to focus the beam on the AGPM like a primary (and secondary) mirror on a telescope. We also use smaller gold coated flat mirrors for beam folding. The entrance pupil and the Lyot stop are iris diaphragms.

### 3.2 Coronagraphic phase mask mount

The AGPM will be placed on a 3-axis mount controlled with actuators. The mount was selected to prevent vignetting of both collimated beams on each side of the AGPM which is typically 10 mm in diameter (see Sect. 4., optical bench layout).

### 3.3 Infrared Camera

The sensor is an infrared InSb camera cooled to 77 K operating at wavelengths ranging from 1.5 to 5 µm. The resolution of 640×512 and the pixel pitch of 15 µm enable to properly sample the Point Spread Function (PSF) of the system in the near-infrared. On VODCA, the camera is used without front lenses to avoid introducing any additional aberrations. The beam is focused on the detector by the parabolas.

### 3.4 Source

One of the requirements of the optical test bench is the possibility to work with broad infrared spectrum from 1 to 5 µm. The source we use is a supercontinuum laser from Le Verre Fluoré with a wavelength coverage fitting the requirements except for the last few hundred nanometers of the spectrum (see Fig. 2). On this bandwidth, another laser diode will be used.

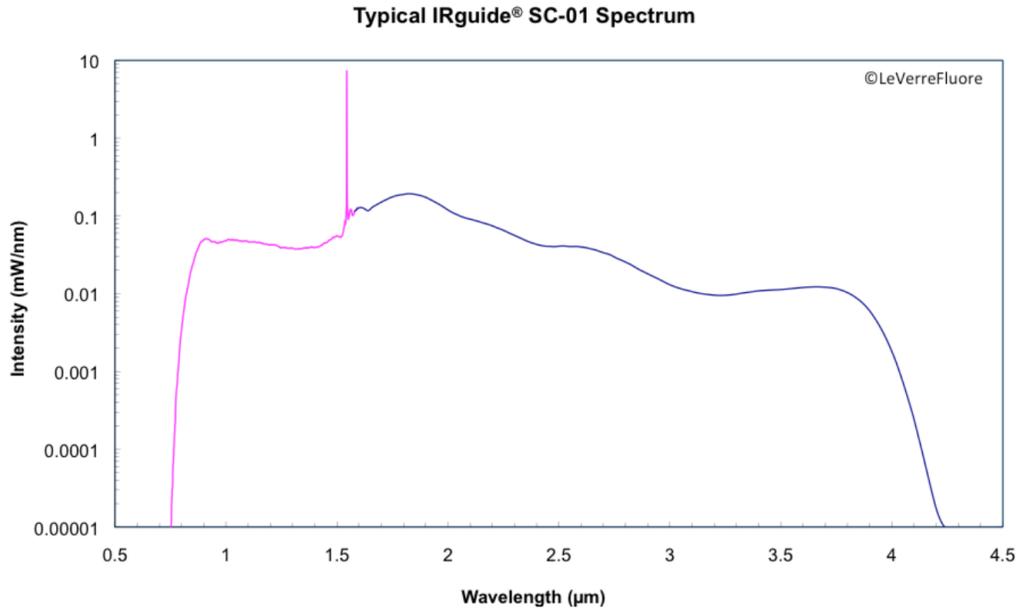

Figure 2. Intensity spectrum of the supercontinuum laser source. © Le Verre Fluoré.

The supercontinuum laser consists of a nanosecond pulsed laser injected into a specific optical fiber. Non-linear optical effects in the fiber allow to broaden the spectrum of the laser. The temporal coherence becomes lower by definition whereas the spatial coherence remains high.

A supercontinuum laser is an excellent source for the tests of achromatic AGPMs. It has the advantage of a halogen lamp, which is a broadband spectrum light. The single mode fiber output provides a spectral density of 10 µW/nm at least, which is not achievable with a halogen lamp on a same numerical aperture (0.23 for the supercontinuum laser source). These specifications guarantee a stable and broadband infrared source on VODCA.

## 4. OPTICAL BENCH LAYOUT

The optical layout of VODCA is shown on Fig 3. A diaphragm is used after the fiber output for two reasons. The starlight cancellation performance of an AGPM is highly dependent on the wavefront quality of the beam. The source is a laser and the output mode of the fiber is the fundamental one (TEM00) which implies a Gaussian form of the amplitude. The first purpose of this diaphragm is to only transmit the central part of the Gaussian which is closer to a uniform amplitude.

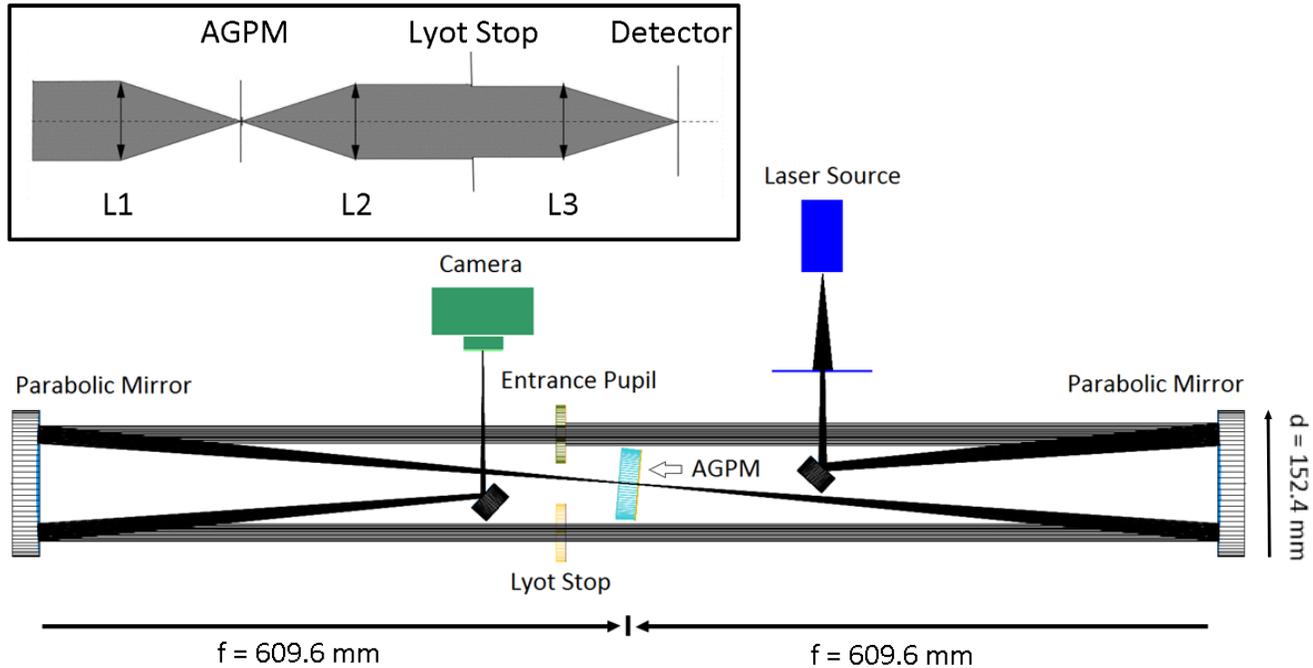

Figure 3. Top: a standard coronagraphic experiment with lenses. Bottom: the Vortex Optical Demonstrator for Coronagraphic Applications (VODCA) setup using two parabolic and two flat mirrors. Designed with the ASAP® software.

Secondly, since the effective focal length of the two parabolic mirrors is fixed, the diaphragm allows the modification of the beam size and by extension the F-number $F\# = f/D$, on the AGPM. In order to fit with the standards on the large telescopes and several infrared high contrast imaging instruments, we set an $F\# = 30$ on the mask. It implies setting the diameter of the entrance pupil of the layout to 20.32 mm. This diameter can be increased up to 30.48 mm to produce a minimum $F\# = 20$ on the bench.

Secondly, since the effective focal length of the two parabolic mirrors is fixed, the diaphragm allows the modification of the beam size and by extension the F-number on the AGPM. In order to fit with the standards on the large telescopes and several infrared high contrast imaging instruments, we set an F-number of 30 on the mask. It implies setting the diameter of the entrance pupil of the layout to 20.32 mm. This diameter can be increased up to 30.48 mm to produce a minimum F-number of 20 on the bench.

If more power is needed on the bench, we could consider using an off-axis parabola to collimate the beam, which numerical aperture would fit the numerical aperture of the fiber in order to keep all of the light emitted by the supercontinuum laser but in this case the beam will present a Gaussian profile. A compromise has to be found between power and uniformity of the wavefront amplitude. It will depend on transmission efficiency on the bench. The first tests performed without an AGPM show a loss of 90% between the output fiber power and the beam power on the camera. As expected, almost all of the losses are due to the first iris diaphragm and the entrance pupil. Based on images of non-coronagraphic point spread function (PSF) which is distributed on 100×100 pixels. It appears that a power of 1 µW on the camera is sufficient to reach a dynamic of $10^4$ with an exposure time of 0.160 ms.

The spectral density of the source is 10 µW/nm and we will use a wavelength range of, at least, 400 nm (K band from 2 to 2.4 µm). Considering these measurements and the power delivered by the source, we will not need to degrade the amplitude uniformity in order to increase the light power on the bench.

As we will use spectral filters and neutral densities, we also intend to implement a setup to filter out aberrations that may be introduced by these components. To do so, the beam at the output of the fiber will be collimated and sent through the filters, and then the beam will be injected again in a single-mode fiber to get rid of all aberrations introduced by the filter. Finally, at the entrance of the coronagraphic part of the optical bench, the wavefront will be flat.

## 5. SIGNAL-TO-NOISE RATIO & CONTRAST

The signal-to-noise ratio (SNR) is a useful metrics to characterize the quality of the test bench. We considered the specifications of the supercontinuum laser source, the infrared camera and a pessimistic estimation of the loss introduced on the optical path by the optics. In the following lines, we define the maximum contrast as the contrast reachable for an SNR of 10.

The L band (centered on 3.8 µm) is the longest infrared wavelengths we will use on VODCA with our infrared supercontinuum light source. The laser source is somewhat less powerful in this wavelength range since it is on the edge of its available spectrum. The infrared background is also becoming more important in this range (assuming a 300 K black body emission). Considering that the L band combines the worst conditions (disregarding the M band, which will be covered by a dedicated light source), we will take it as example for the maximum contrast estimation we can achieve on VODCA.

The detector being cooled to 77 K, its intrinsic thermal background will not contribute significantly to the overall noise budget. Taking into account the estimated read-out noise of the detector (1000 electrons rms) and the environmental background emission, we expect to achieve an SNR of about $10^4$ on the peak in 1 sec of total integration time (adding up individual frames of a few tens of msec each). Inserting the AGPM, we would then still obtain an SNR of 10 on the peak if the peak rejection amounts to $10^6$. This is amply sufficient to test the AGPM performance, which should provide a peak rejection of about $10^3$. It would then even be possible to measure the coronagraphic PSF profile with a dynamic range of about $10^3$.

This level of performance is good enough for our applications, but could still be increased if needed, e.g., by increasing the total integration time. The addition of a reflective baffle (currently under study) at the entrance of the camera could further reduce the environmental background noise to get a better contrast.

## 6. EXPERIMENT OF PRE- AND POST- CORONAGRAPHIC CONCEPTS

The development of VODCA is also the opportunity to test pre-coronagraphic concepts like the use of an apodizer before the AGPM or post coronagraphic concepts like the reflective Lyot stop

### 6.1 Optimal apodization

As all focal plane phase mask coronagraphs, the AGPM only provides its best performance on a non-obscured input pupil. However, all ground-based telescopes feature a central obscuration due to the presence of the secondary mirror, and spider arms to hold the secondary. It has recently been proposed to use a ring-type apodization in the entrance pupil to remove the additional light inside the Lyot stop created by the diffraction of the secondary mirror structure [6].

We plan to test this apodized solution on VODCA by creating a central obscuration in the input pupil and introducing an amplitude apodization in that same pupil. This would be a most-welcome technical demonstration before installing this kind of solution on an actual instrument.

**6.2 Reflective Lyot stop**

The starlight cancellation of a coronagraph is highly dependent on the wavefront aberrations, especially tip and tilt, which are low order aberrations with high amplitudes. Most of the tip/tilt aberrations are corrected by a first adaptive optics loop on a telescope, but some non-common path aberrations can still occur between this correction and the coronagraph.

The Lyot stop is an optical stop placed in a pupil plane after the coronagraphic mask, slightly smaller than the beam size. Its purpose is to block the starlight diffracted by the coronagraph.

The concept we want to test on VODCA is to analyze the blocked light by using a reflective Lyot stop and a camera to measure with high sensitivity the amplitude of the remaining tip and tilt as proposed by Singh et al [4]. Then, the aberrations will be corrected by a deformable mirror introduced on the optical bench.

## 7. CONCLUSIONS

In this manuscript we have detailed the development of VODCA: the coronagraphic optical test bench in development at the University of Liège. The achromatic optical bench design has been explained and we briefly described the optics components.

VODCA is built to assess the performance of SGVCs such as the AGPM, before integrating them on large telescope instruments, but it is also the opportunity to test coronagraphic theories. A reflective Lyot stop and a deformable mirror will correct tip and tilt aberrations on the phase mask. VODCA will help to the validation of coronagraphic concepts which aim to improve the quality of direct imaging of closer and fainter star companions.

## ACKNOWLEDGMENTS

The research leading to these results has received funding from the European Research Council under the European Union's Seventh Framework Programme (ERC Grant Agreement n.337569) and from the French Community of Belgium through an ARC grant for Concerted Research Actions.